\documentclass[twocolumn]{aastex631}

\usepackage[T1]{fontenc}
\usepackage{amsmath}
\usepackage{booktabs}
\usepackage{hyperref}

\newcommand{\sm}{$M_\sun$}

\providecommand{\Unit}[1]{\,\mathrm{#1}}

\shorttitle{Simulating ``Fermi Bubble'' in the Circinus Galaxy}
\shortauthors{Xie et al.}
\defcitealias{mingo2012}{M12}

\begin{document}

\title{Simulating the Formation of the Young ``Fermi Bubbles'' in the Circinus Galaxy}

\correspondingauthor{Fulai Guo}
\email{fulai@shao.ac.cn}

\author[0000-0001-5624-6008]{Shaokun Xie}
\affiliation{Shanghai Astronomical Observatory, Chinese Academy of Sciences, 80 Nandan Road, Shanghai 200030, People’s Republic of China}
\affiliation{University of Chinese Academy of Sciences, 19A Yuquan Road, Beijing 100049, People’s Republic of China}

\author[0000-0003-1474-8899]{Fulai Guo}
\affiliation{Shanghai Astronomical Observatory, Chinese Academy of Sciences, 80 Nandan Road, Shanghai 200030, People’s Republic of China}
\affiliation{University of Chinese Academy of Sciences, 19A Yuquan Road, Beijing 100049, People’s Republic of China}
\affiliation{Tianfu Cosmic Ray Research Center, 610000 Chengdu, Sichuan, People’s Republic of China}

\author[0000-0002-6563-7438]{Ruiyu Zhang}
\affiliation{School of Physics, Henan Normal University, Xinxiang 453000, People’s Republic of China}
\affiliation{Center for Theoretical Physics, Henan Normal University, Xinxiang 453000, People’s Republic of China}

\author[0000-0001-5649-938X]{B. Mingo}
\affiliation{Centre for Astrophysics Research, University of Hertfordshire, College Lane, Hatfield AL10 9AB, UK}

\author[0000-0003-3922-5007]{Fangzheng Shi}
\affiliation{Shanghai Astronomical Observatory, Chinese Academy of Sciences, 80 Nandan Road, Shanghai 200030, People’s Republic of China}
\affiliation{INAF - Osservatorio Astronomico di Roma, Via Frascati 33, 00078, Monte Porzio Catone (Roma), Italy}

\author[0000-0001-6239-3821]{Jiang-Tao Li}
\affiliation{Purple Mountain Observatory, Chinese Academy of Sciences, 10 Yuanhua Road, Nanjing 210023, People’s Republic of China}

\begin{abstract}
The Fermi and eROSITA bubbles in the Milky Way represent an archetypal case of galactic nucleus feedback, yet their origin remains highly debated. Here we use hydrodynamic simulations to investigate the formation of the ``Fermi bubbles'' in the nearby Circinus galaxy, a pair of kpc-scaled elliptical bubbles seen in both radio and X-ray observations. We find that a pair of active galactic nucleus (AGN) jets drive forward shocks in the circumgalactic medium, and after evolving for $\sim 0.95\Unit{Myr}$, the shock-delineated bubble pair roughly matches the observed Circinus bubbles in size and morphology. Our mock X-ray image and spectrum reproduce the observed edge-brightened X-ray surface brightness distribution and spectrum quite well, and suggest that non-thermal emissions from the jet ejecta also contribute substantially to radio and X-ray emissions from the inner ``hotspot'' region. We further show that AGN winds tend to produce more spherical bubbles with a wider base near the galactic plane, inconsistent with observations. The hotspot emissions and the misalignment between the galaxy rotational axis and the bubble's axis argue against a starburst wind origin. Our study thus corroborates the AGN jet-shock model for the origin of both the Circinus bubbles and the Fermi bubbles, and suggests that AGN jet feedback may be a common origin of extended gaseous bubbles in regular disk galaxies, potentially playing an important role in their evolution.
\end{abstract}

\section{Introduction\label{sec:intro}}
Active galactic nucleus (AGN) feedback is widely recognized as a key physical process affecting the evolution of massive early-type galaxies and galaxy clusters, primarily by heating gas in the circumgalactic medium (CGM) and suppressing gas accretion from the intergalactic medium \citep{mcnamara:nulsen:2007, mcnamara:nulsen::2012, fabian2012,2020NewAR..8801539H,donahue22,bourne23}. One of the most compelling evidence for AGN feedback comes from observations of numerous X-ray cavities in galaxy clusters and elliptical galaxies, which are often associated with radio jets and spatially coincident with radio lobes (e.g., \citealt{Boehringer1993, fabian2002, croston2011, birzan2020, sonkamble2024}). In contrast, stellar feedback has traditionally been considered as the dominant regulatory mechanism in the evolution of late-type galaxies, partly due to the relatively scarcity of clear galaxy-scale AGN feedback signatures. However, with improving instrumental sensitivity, growing evidence suggests that AGN activity is also quite common and probably plays an important role in spiral galaxies \citep{cecil2000, veilleux1994, li2019, pietsch1998, irwin2003, sebastian2019, li2022, carilli1992, croston2008, heesen2011, mingo2012, zeng2023}. Among the most striking examples are the Fermi bubbles, discovered in the Milky Way in the gamma ray band ($1 \lesssim E_\mathrm{\gamma} \lesssim 200$ GeV), with a spatial size of $\sim 10\Unit{kpc}$ \citep{su2010}. The spatially coincident microwave haze \citep{finkbeiner2004, dobler2008} and giant X-ray eROSITA bubbles \citep{predehl2020} further point to past episodes of nuclear activities - possibly AGN feedback - in the Milky Way. 

In the nearby Seyfert 2 galaxy Circinus ($\sim 4.2\Unit{Mpc}$, \citet{tully2009, zschaechner2016}), Chandra observations revealed a pair of $\sim 3\Unit{kpc}$ bubbles in the $0.4-5.0\Unit{keV}$ band \citep[hereafter M12]{mingo2012}, both of which show a shell-like, edge-brightened X-ray morphology and a corresponding edge-brightened radio structure \citep{elmouttie1998}. The Circinus bubbles are strongly reminiscent of the Fermi bubbles in the Milky Way, and have a slightly smaller size potentially in a younger evolution stage.

Although the Circinus galaxy is a starburst system with a star formation rate of 3-8 \sm/yr \citep{for2012}, the Circinus bubbles are unlikely to be driven by the starburst wind, which could not easily explain the enhanced radio and X-ray emissions from the ``hotspot'' regions near the bubble center (see \citetalias{mingo2012}). \citetalias{mingo2012} also argued that the radio polarization level in the west bubble and the variation of the radio spectral index across the bubbles disfavor the starburst wind model. In addition, we find that the west bubble is misaligned by $\sim 37\degr$ with respect to the galactic rotation axis, which strongly disfavors a stellar-feedback origin, as random supernova explosions in the galactic nucleus tend to produce outflows aligned roughly perpendicular to the disk over time. Intriguingly, the deviation of $\sim 37\degr$ is in excellent agreement with the $40\degr$ tilt of the nuclear accretion disk proposed by \citet{stalevski2017} to account for the peculiar mid-infrared morphology of the Circinus core -- a geometry subsequently corroborated by \citet{stalevski2019, stalevski2023}. This striking alignment provides compelling evidence linking the bubbles directly to AGN activity.
 
The Circinus galaxy hosts a supermassive black hole of $1.7\times 10^6\;M_\odot$ \citep{Greenhill2003}, comparable to that of Sgr A* in the Milky Way. Its AGN exhibits a bolometric luminosity of $1.7\times 10^{10}\;L_\odot$ \citep{maiolino1998}, sufficient to photo-ionize the gas in the nuclear region and produce an ionization cone that spatially coincides with the base of the west bubble \citep{marconi1994, elmouttie1998b, wilson2000, smith:wilson:2001}. However, as argued by \citetalias{mingo2012}, AGN photoionization alone cannot explain the observed $\sim 3\Unit{kpc}$ bubbles: given the high density of the surrounding circumnuclear environment, the ionization radius is limited to only $\sim 700\Unit{pc}$. This shortfall rules out pure radiative driving and, together with the enhanced hotspot emissions, strongly supports a mechanical origin -- most plausibly by an AGN jet pair -- for the large-scale bubble structure.
 
AGN jets have been proposed to explain the formation of Fermi bubbles in the Milky Way, whose origin remains elusive (see recent reviews by \citealt{yang18} and \citealt{sarkar24}). They are often interpreted as the ejecta bubbles due to AGN feedback (e.g., \citealt{guo2012a, yang2012a,mou2014,yang2022}) or stellar feedback (e.g., \citealt{crocker15,sarkar15}). In contrast, \citet{zhang2020} interpret the Fermi bubbles as the shock-enclosed CGM bubbles driven by a past AGN jet event, while the faded jet ejecta are hidden inside the observed Fermi bubbles. This AGN jet-shock model naturally explains the X-shaped biconical X-ray structure coincident with the base of Fermi bubbles \citep{snowden1997, bland-hawthorn:cohen:2003, keshet:gurwich:2017}, and cosmic-ray electrons (CRes) may be accelerated in-situ by the forward shock (see observational evidence in \citealt{li2019}), thus solving the short cooling time problem associated with previous AGN-jet models, which usually assume that CRes are accelerated at the jet base and then transported to the bubble interior \citep{guo2012a, yang2012a, yang2022}. The Circinus bubbles show an edge-brightened shell-like structure, which is a clear indication of forward shock, and contain an emission-enhanced ``hotspot'' region (\citetalias{mingo2012}), which can be interpreted as the bubble filled with the jet ejecta (the ``ejecta bubble''). The striking morphological and energetic similarities between the Circinus and Fermi bubbles strongly motivate us to test the AGN jet-shock scenario for the former, which are younger and contain an inner not-yet-faded emitting ejecta bubble.
 
In this study, we used hydrodynamic simulations to investigate whether the X-ray bubbles observed in the nearby Circinus galaxy can be explained by forward shocks driven by a recent AGN jet event. In addition to bubble morphology, we also investigated whether the observed X-ray emission mainly originates from the shock-compressed CGM shell and compare our mock observation results directly with the X-ray observations by \citetalias{mingo2012}. The remainder of the paper is organized as follows. In Section \ref{sec:method}, we describe the details of our simulation setup, the Circinus galaxy model, and the jet implementation method. The results of our fiducial run and the comparison with X-ray observations are presented in Section \ref{sec:results}. Then in Section \ref{sec:discussion} we discuss the uncertainties in some model parameters and investigate the AGN wind model with two additional runs. We summarize our main results in Section \ref{sec:summary}. 

\section{Methods\label{sec:method}}
\subsection{Simulation Setup\label{sec:code}}
Our simulations are performed under a 3-dimensional (3D) hydrodynamic framework with the publicly available code {\tt PLUTO} \citep{mignone2007}. We use the piecewise parabolic method ({\tt PPM}) for reconstruction, the third-order TVD Runge Kutta scheme for time integration, and the {\tt HLLC} Riemann solver to solve the hydrodynamic equations.

The eastern Circinus bubble is heavily obscured by the gaseous disk of Circinus, and therefore we focus on modeling the evolution of a one-sided jet and perform quantitative comparisons with the observations of the less-obscured western Circinus bubble. To minimize computational cost, the simulation domain is chosen to be only slightly larger than the western bubble, whose edge corresponds to the jet-driven forward shock. Adopting the distance to the Circinus galaxy of $D = 4.2$ Mpc \citep{Karachentsev2013} and an inclination angle of the Circinus disk of $\sim 65 \degr$ derived from HI kinematics \citep{freeman1977}, and further assuming no additional tilt of the bubble axis along the line of sight, the de-projected physical size of the western bubble along the jet direction is $2.7\Unit{kpc}$. This value corresponds to the length of the major axis of the western Circinus bubble that our simulations aim to reproduce.

Our simulations are performed in 3D Cartesian coordinates that span the physical domain $x,y \in [-1, 1]\Unit{kpc}$ and $z \in [0, 3.3]\Unit{kpc}$, centered on the galactic nucleus. We use a uniform grid of $400\times400\times660$ cells. The lower boundary at $z=0$ corresponds to the galactic mid-plane and is treated with reflective boundary conditions. All other boundaries are initialized with hydrostatic equilibrium and the physical boundary properties are thereafter held fixed throughout our simulations. This boundary condition substantially reduces spurious numerical errors that often arise when using standard outflow conditions with a finite-volume algorithm. It is also physically well-justified: the initial CGM is assumed to be in hydrostatic equilibrium, and the jet-driven shocks and perturbations have not yet reached the outer boundaries at the end of our simulations.

\subsection{The Circinus Galaxy Model\label{sec:gal_model}}
The Circinus galaxy is a disk galaxy, and its gravitational potential in our simulation domain comes mainly from two sources: a stellar bulge and a stellar disk
\begin{equation}
	\Phi = \Phi_\mathrm{disk} + \Phi_\mathrm{bulge} ~{.}
\end{equation}
Here we neglect the gravitational potential of the dark matter halo, as its contribution is minor in our simulation domain, and the observational constraints on its properties are quite limited. The Milky Way has a dark matter halo with a mass comparable to that of Circinus, and we adopt the Milky Way dark matter distribution \citep{mcmillan2017} to model that of Circinus, confirming that the stellar bulge and disk dominate the gravitational acceleration of Circinus within $r \sim 7$ kpc, exceeding our simulation domain. 

We assume that the gravitational potentials of the stellar disk and bulge follow the profiles of a Miyamoto-Nagai disk \citep{miyamoto1975} and a spherical Jaffe stellar bulge \citep{jaffe1983}, respectively:
\begin{eqnarray}
	\Phi_{\mathrm{disc}} =& -\frac{G M_{\mathrm{d}}}{\sqrt{R^{2}+\left(a+\sqrt{z^{2}+b^{2}}\right)^{2}}} ~{,}\\
	\Phi_{\text {bulge}} =&- \frac{G M_{\text {b}}}{r_\mathrm j}\ln\left(\frac{r}{r+r_{\mathrm{j}}}\right)~{,}
\end{eqnarray}
where R is the galactocentric radius in the galactic plane and $r = \sqrt{R^2+z^2}$ is the spatial distance to the galactic center. The stellar bulge mass $M_\mathrm b$ and the scale length $a$, $b$, $r_\mathrm j$ are the free model parameters. The mass of the stellar disk is determined by subtracting $M_\mathrm b$ from the total stellar mass of the Circinus galaxy $M_{*} \sim 9.5\times 10^{10}$ \sm~\citep{for2012}.

\begin{figure}[ht!]
\plotone{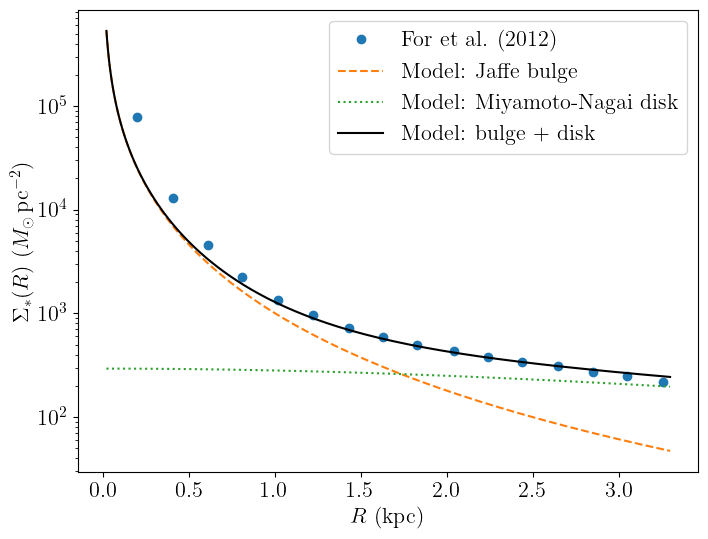}
\caption{Stellar mass surface density profile of our Circinus model (solid line) compared with the observational measurements (points; \citealt{for2012}). \label{fig:stellar_model}}
\end{figure}

The mass density profiles of the stellar disk and bulge in our model are, respectively,
\begin{eqnarray}
&\rho_{\mathrm{disc}} = \frac{b^2 M_{\mathrm{d}}}{4\pi} \frac{aR^2+\left(a+3\sqrt{z^{2}+b^{2}}\right)\left(a+\sqrt{z^{2}+b^{2}}\right)^{2}}{\left[ R^2 + \left(a+\sqrt{z^2+b^2}\right)^2\right]^{5/2}\left(z^2+b^2\right)^{3/2}}~{,}\\
&\rho_{\text {bulge}} = \frac{M_{\text {b}}}{4\pi r^3_\mathrm j}\frac{r^4_\mathrm j}{r^2\left(r+r_\mathrm j\right)^2}~{.}
\end{eqnarray}
With the above profiles, we calculate the stellar mass surface density profile, and by fitting it with the observed stellar mass surface density profile \citep{for2012}, we derive the best-fitting model parameters, which are listed in Table \ref{tab:gal}. Fig. \ref{fig:stellar_model} shows our best-fit stellar mass surface density profile and the contributions from the stellar disk and bulge.

\begin{table}
\caption{Parameters in our Circinus model\label{tab:gal}}
\begin{tabular}{lcr}
\toprule
Halo gas temperature & $T_\mathrm h$ & 0.3 keV \\
Stellar bulge mass & $M_\mathrm{b}$ & $3\times 10^{10}\;\mathrm M_\sun$\\
Scale radius of stellar bulge & $r_\mathrm j$ & 0.6 kpc \\
Stellar disk mass & $M_\mathrm{d}$ &  $6.5\times 10^{10}\;\mathrm M_\sun$ \\
Scale radius of stellar disk & $a$ & 5.9 kpc \\
Scale height of stellar disk & $b$ & 0.26 kpc \\
\bottomrule
\end{tabular}
\end{table}

At the beginning of our simulations, the CGM of Circinus is assumed to be isothermal with temperature $T_\mathrm h = 0.3$ keV and in hydrostatic equilibrium in our modeled axisymmetric gravitational potential. The CGM density distribution can be derived as  
\begin{equation}
	\rho(R, z) = \rho(R_0,z_0) \exp\left[\frac{\Phi(R_0,z_0) - \Phi(R,z)}{k_\mathrm{B}T_\mathrm{h}/(\mu m_\mathrm{p})}\right] ~{,}
\end{equation}
where $k_\mathrm{B}$ is the Boltzmann constant, $\mu = 0.61$ is the mean molecular mass, and $m_\mathrm p$ is the atomic mass unit. The distribution of the gas density is normalized by $\rho(R_0, z_0) = 0.03\;m_\mathrm{p}\Unit{cm^{-3}}$ at the reference location $(R_0, z_0)=(0,1\Unit{kpc})$. 

\subsection{Jet Implementation\label{sec:jet}}
A real AGN jet is launched near the black hole's event horizon, which is far below our grid resolution, and therefore cannot be self-consistently modeled in our simulations. Here we instead inject a one-sided jet along the $z$ axis into our simulation hemisphere through a cylindrical nozzle located at the galactic nucleus. When the jet is active, the fluid variables within this nozzle are fixed to be the jet properties and do not evolve with time. The jet is then injected into our simulation domain through the upper surface of the nozzle (internal boundary in {\tt PLUTO}; see \citealt{Mukherjee20}). The nozzle has a radius of 15 pc (6 grid cells), wide enough to resolve Kelvin-Helmholtz instabilities, and a vertical length of 175 pc along the $z$ axis from the galactic center, which ensures that the injection region is placed outside the complex central environment. This setup also allows us to adopt a sub-relativistic jet velocity which substantially increases the associated Courant–Friedrichs–Lewy timestep and reduces the computational cost compared to a fully relativistic treatment. The jet is instantly turned off once the jet-driven forward shock reaches a prescribed height ($z=2.2\Unit{kpc}$), and the cylindrical jet nozzle reverts to normal evolution afterwards. The simulation is then continued without further jet injection until the forward shock reaches the observed size of the western Circinus bubble ($z = 2.7\Unit{kpc}$).

\begin{deluxetable*}{lccccccccccc}
\tablecaption{Parameters and derived properties of AGN jets or winds in our simulations\label{tab:runs}}
\tablehead{
  & \colhead{$\theta^*$} & \colhead{$\eta^*$} & \colhead{$\rho_\mathrm j$} & \colhead{$\kappa^*$} & \colhead{$e_\mathrm j$} & \colhead{$v_\mathrm j^*$} & \colhead{$t_\mathrm j$} & \colhead{$P_\mathrm j$} & \colhead{$E_\mathrm{j}$} & \colhead{$t_\mathrm{bub}$}\\
 \colhead{Run} & \colhead{($\degr$)} &  & \colhead{($10^{-26}\;\mathrm{g\,cm^{-3}}$)} & & \colhead{($10^{-10}\;\mathrm{erg\,cm^{-3}}$)} & \colhead{($c$)} & \colhead{(Myr)} &  \colhead{($10^{42}\;\mathrm{erg\,s^{-1}}$)} & \colhead{($10^{55}\;\mathrm{erg}$)} & \colhead{(Myr)} \\
}
\startdata
\texttt{jet}         & 0   & 0.01 &  0.05  & 15 & 3.5 & 0.25 & 0.5   & 1.5   & 2.3   & 0.95 \\
\texttt{wind60} & 60 & 1      & 5        & 10 & 2.4 & 0.03 & 1.3   & 15.2 & 62.9 & 1.42 \\
\texttt{wind30} & 30 & 1      &  5       & 10 & 2.4 & 0.03 & 1.09 & 4.08 & 14.03 & 1.18 
\enddata
\tablecomments{
The variables with an asterisk are our model parameters of AGN jets or winds, while the others are properties derived from these parameters. $\theta$ - half-opening angle,  $\eta$ - density ratio between jet and ambient gas, $\rho_\mathrm j$ - jet density, $\kappa$ - pressure ratio between jet and ambient gas, $e_\mathrm j$ - energy density of jet, $v_\mathrm j$ - jet velocity, $t_\mathrm j$ - jet duration, $P_\mathrm j$ - power of the jet pair, $E_\mathrm{j}$ - total energy injected by the jet pair, $t_\mathrm{bub}$ - current bubble age.
}
\end{deluxetable*}

We characterize the jet properties with three model parameters: density contrast ($\eta = \rho_j/\rho_0$), pressure contrast ($\kappa = p_j/p_0$) and jet velocity $v_\mathrm j$. Gas mass, momentum and energy are injected into the computational domain through the upper surface of the jet nozzle, and the instantaneous jet power may be calculated as

\begin{eqnarray} \label{eqn:jet}
	P_\mathrm j &=& P_\mathrm{kin} + P_\mathrm{th} \nonumber \\
	&=& \dot{m_\mathrm j} \left(v^2_\mathrm j /2 + e_\mathrm j \right) \nonumber \\
	&=& \eta \rho_0 v_\mathrm j  \pi r^2_\mathrm j \left[v^2_\mathrm j /2 + \kappa p_0/(\gamma -1) \right] ~{,}
\end{eqnarray}
where $r_\mathrm{j}$ is the radius of the jet nozzle, $\gamma=5/3$ is the adiabatic index of the ideal gas, and $\rho_0$, $p_0$ are the density and thermal pressure of the ambient gas at the reference point $(R_0, z_0) = (0, 1\Unit{kpc})$, respectively. In Table \ref{tab:runs}, we list the jet parameters and properties (including the jet power $P_\mathrm{j}$ and total injected energy $E_\mathrm{j}$) in a representative simulation (run \texttt{jet}), which reproduces the observed properties of the west bubble reasonably well.

\section{results\label{sec:results}}
In this section, we present the results of our representative jet simulation (run \texttt{jet} shown in Table \ref{tab:runs}), which reproduces the key observational properties of the western Circinus bubble in the Chandra $0.4-5.0$ keV band (\citealt{mingo2012}) quite well. The jet in this run is highly under-dense ($\eta=0.01$), over-pressured ($\kappa=15$) relative to the ambient medium, and kinetic-energy-dominated with a kinetic fraction of $\sim 96\%$ in $P_\mathrm{j}=1.5\times 10^{42}\Unit{erg/s}$ (the total power of the jet pair). The jet remains active for $0.5\Unit{Myr}$ and then is turned off afterwards. The forward shock continues to propagate outward, and at $t_\mathrm{bub} \simeq 0.95\Unit{Myr}$, reaches $z=2.7\Unit{kpc}$ along the $z$ axis, the de-projected height of the outer edge of the observed western Circinus bubble.

\begin{figure*}[ht!]
\includegraphics[width=\textwidth]{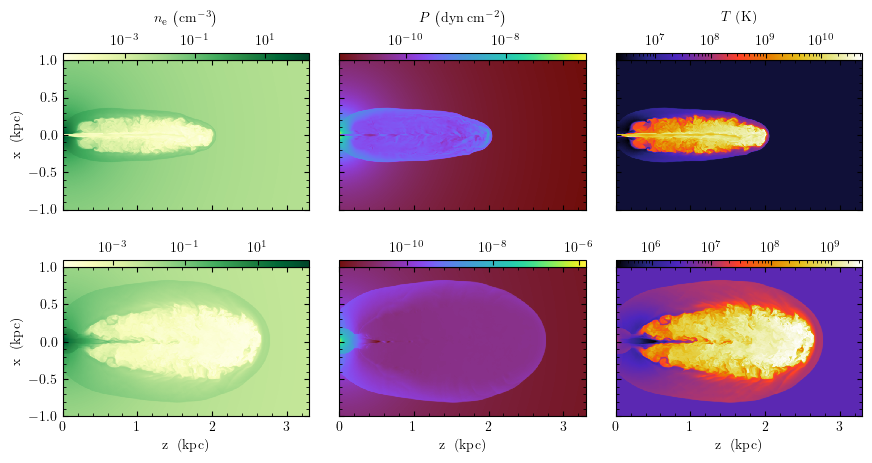}
\caption{Central slices in the $x-z$ plane in our fiducial jet run at two representative epochs: just before the jet termination ($t = 0.46\Unit{Myr}$, upper row), and at the end of the run ($t = 0.95\Unit{Myr}$, lower row) when the forward shock reaches $z \approx 2.7\Unit{kpc}$, forming the observed western bubble. From left to right, the columns show the distributions of electron number density, thermal pressure, and temperature, respectively. The jet evolution here follows the classic evolution pattern for light, over-pressured jets: a hot elongated cocoon during its active phase gradually transitions to a more spherical bubble enclosed by a shocked gaseous shell after the jet switch-off, naturally explaining the observed limb-brightened X-ray bubble.}
\label{fig:2d}
\end{figure*}

\subsection{Formation of the Circinus Bubbles}

In Figure \ref{fig:2d}, we present the central slices of the electron number density, thermal pressure, and temperature distributions at two representative epochs: just before jet termination ($t=0.46\Unit{Myr}$, upper row) and at the end of the simulation ($t=0.95\Unit{Myr}$, lower row). The morphological evolution closely resembles that seen in previous studies of light, over-pressured jets \citep{sutherland2007,bromberg2011,wagner2011}. The jet ejecta form a lobe, which is extremely hot (up to several $10^9\Unit{K}$) and tenuous. In contrast, the outer forward shock heats and compresses the ambient CGM, producing a thin post-shock shell of higher density, pressure, and temperature. While the jet is active, the continuous momentum injection keeps both the lobe and the shocked CGM bubble substantially elongated along the $z$ axis. Once the jet is turned off, the gas pressure in the cocoon  is roughly uniform and drives lateral expansion, causing the entire structure to gradually evolve towards a more spherical shape.

For an optically thin plasma in collisional ionization equilibrium, the volumic X-ray emissivity is a function of the gas temperature $T$ and metallicity $Z$, and is proportional to the gas density squared: $\epsilon(T, Z) = n_\mathrm{e} n_\mathrm{H} \Lambda(Z, T)$. The brightest X-ray emission thus arises overwhelmingly from the densest region in our simulation. The shocked CGM shell reaches electron densities of $n_\mathrm{e}\sim 0.1-1\Unit{cm^{-3}}$, a factor of $\gtrsim 30$ higher than that in the rarefied lobe filled with the jet ejecta ($n_\mathrm{e}\lesssim 3\times 10^{-3}\Unit{cm^{-3}}$). This naturally produces a limb-brightened surface-brightness distribution in the soft X-ray band. We therefore infer that the shell-like X-ray emission of the Circinus bubble originates mostly from the thermally emitting post-shock region, and further explore our simulation results at $t_\mathrm{bub} \simeq 0.95\Unit{Myr}$ in the following two subsections.
  
\subsection{Comparison with X-ray Observations}

\begin{figure}
	\plotone{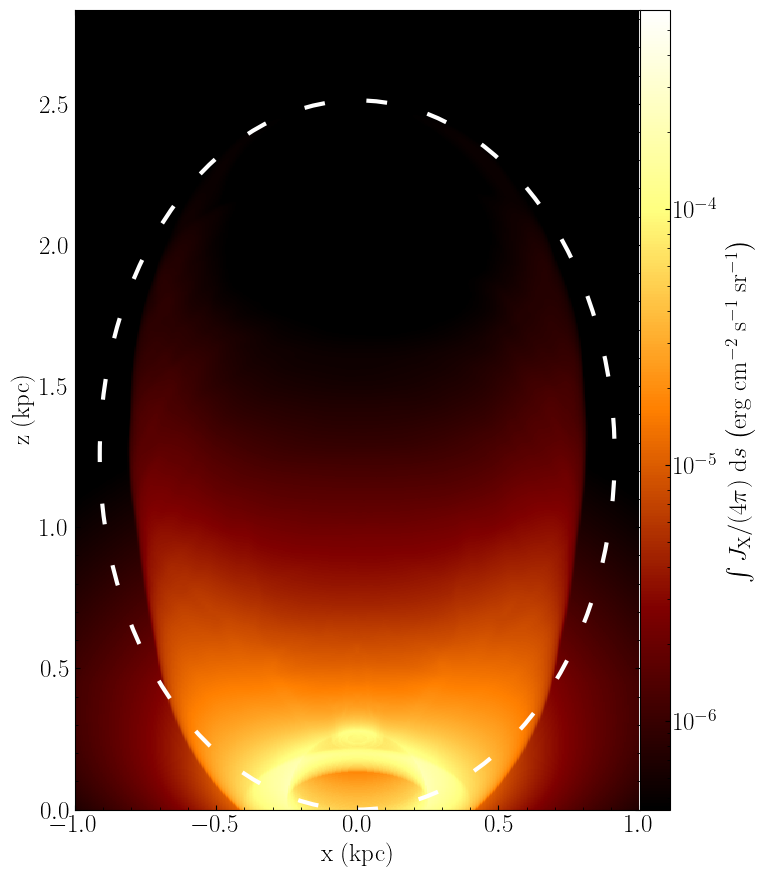}
	\caption{Synthetic 0.4 – 5.0 keV X-ray surface brightness distribution at $t_\mathrm{bub} \simeq 0.95\Unit{Myr}$ in our fiducial run \texttt{jet} (APEC model, $Z = 0.25\;Z_\odot$). The central $0.25\Unit{kpc}$ is masked to suppress the nuclear emission. The dashed ellipse marks the observed outer boundary of the western bubble \citep{mingo2012}. The simulated bubble matches its observed size (bubble major axis $\sim 2.5$ kpc), shape, and limb-brightened morphology reasonably well. \label{fig:xray_map}}
\end{figure}

Fig. \ref{fig:xray_map} shows the synthetic $0.4-5.0\Unit{keV}$ X-ray surface brightness map at $t_\mathrm{bub} \simeq 0.95\Unit{Myr}$ in our fiducial run \texttt{jet}, overlaid with a dashed ellipse that approximates the observed outer boundary of the western Circinus bubble. The surface brightness is calculated by integrating the $0.4-5.0\Unit{keV}$ X-ray emissivity $\epsilon(T, Z)$, divided by $4\pi$ \citep{zhang2020}, along the line of sight at the galaxy's observed inclination angle using the \texttt{yt} astrophysics analysis software \citep{turk2011}, and here $\epsilon(T, Z)$ is computed using the \texttt{APEC} plasma model with fixed metallicity $Z=0.25\,Z_\odot$. To avoid the extremely bright nuclear emission from dominating the map, we mask the central region within $0.25\Unit{kpc}$ of the galaxy center -- a zone beyond the scope of the present study. The limb-brightened bubble in the synthetic map closely matches the size and position of the observed western Circinus bubble, demonstrating that the AGN jet-shock model could successfully reproduce this structure, as elaborated in more detail in the following.

The pronounced limb-brightened appearance arises from two physical effects. First, the post-shock region contains significantly denser plasma, yielding much stronger bremsstrahlung and line emissions than the rarefied bubble interior. Second, the line-of-sight projection causes emissions from the near and far sides of the shell to accumulate along the bubble edge, further enhancing the observed bright rim. Together, these effects naturally produce the sharp, edge-brightened morphology seen in the Chandra data.
        
X-ray emission from the bubbles is subject to foreground absorption, as Circinus is located behind the Galactic plane, and contamination from foreground radiation, while instrument response further affects the detection efficiency. These effects may easily overwhelm the relatively faint signal from extragalactic sources. To properly account for these realistic observing conditions, we use \texttt{pyXSIM}\footnote{\url{https://hea-www.cfa.harvard.edu/~jzuhone/pyxsim/}} to convert the simulated bubble into an X-ray photon list and \texttt{SOXS}\footnote{\url{https://hea-www.cfa.harvard.edu/soxs/index.html}} to generate the mock Chandra ACIS-S observation. The photon list is produced with the same \texttt{APEC} emission model and parameters as in Fig. \ref{fig:xray_map}, assuming a distance of $4.2\Unit{Mpc}$ and an exposure time of $300\Unit{ks}$. Galactic foreground absorption is applied using a hydrogen column density of $N_\mathrm{H} = 5.2 \times 10^{21}\Unit{cm^{-2}}$, adopted from \texttt{FTOOLS} in the direction of the Circinus galaxy \citep{blackburn1999}.

\begin{figure*}
	\centering
    \includegraphics[height=7cm]{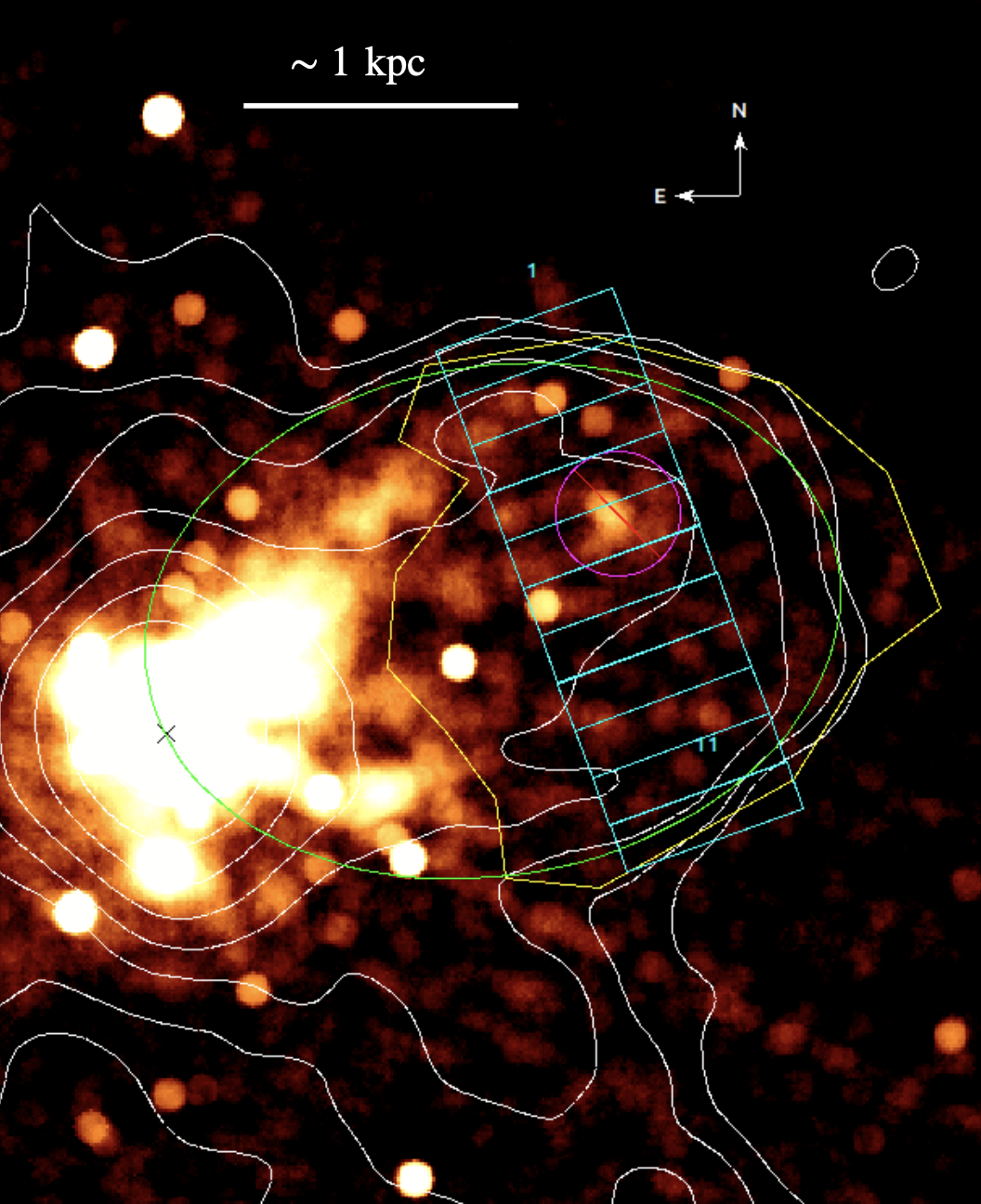}
     \hspace{0.1cm}
    \includegraphics[height=7cm]{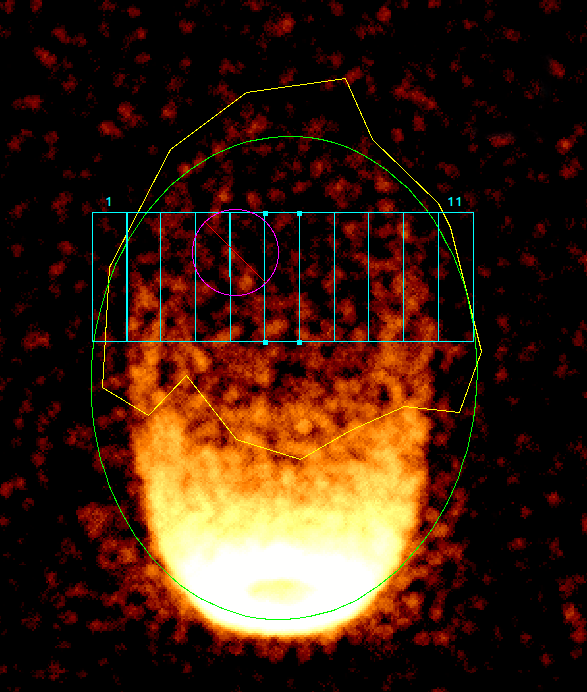}
    \caption{Left: the observed Chandra 0.4 – 5.0 keV image of the Circinus western bubble with overlaid ATCA 13 cm radio contours (white; \citetalias{mingo2012}). Right: the mock Chandra observation generated with \texttt{pyXSIM} + \texttt{SOXS} from our fiducial simulation at $t_\mathrm{bub} \simeq 0.95\Unit{Myr}$, including the effects of the Galactic foreground emission and absorption, instrument background, and realistic detector response (300 ks exposure). The green oval, which roughly shows the outer boundary of the bubbles, has the identical physical size in both panels. Our simulation faithfully reproduces the observed bubble extent, shell thickness, and surface-brightness contrast.}
	\label{fig:mock_xray}
\end{figure*}

\begin{figure}
	\plotone{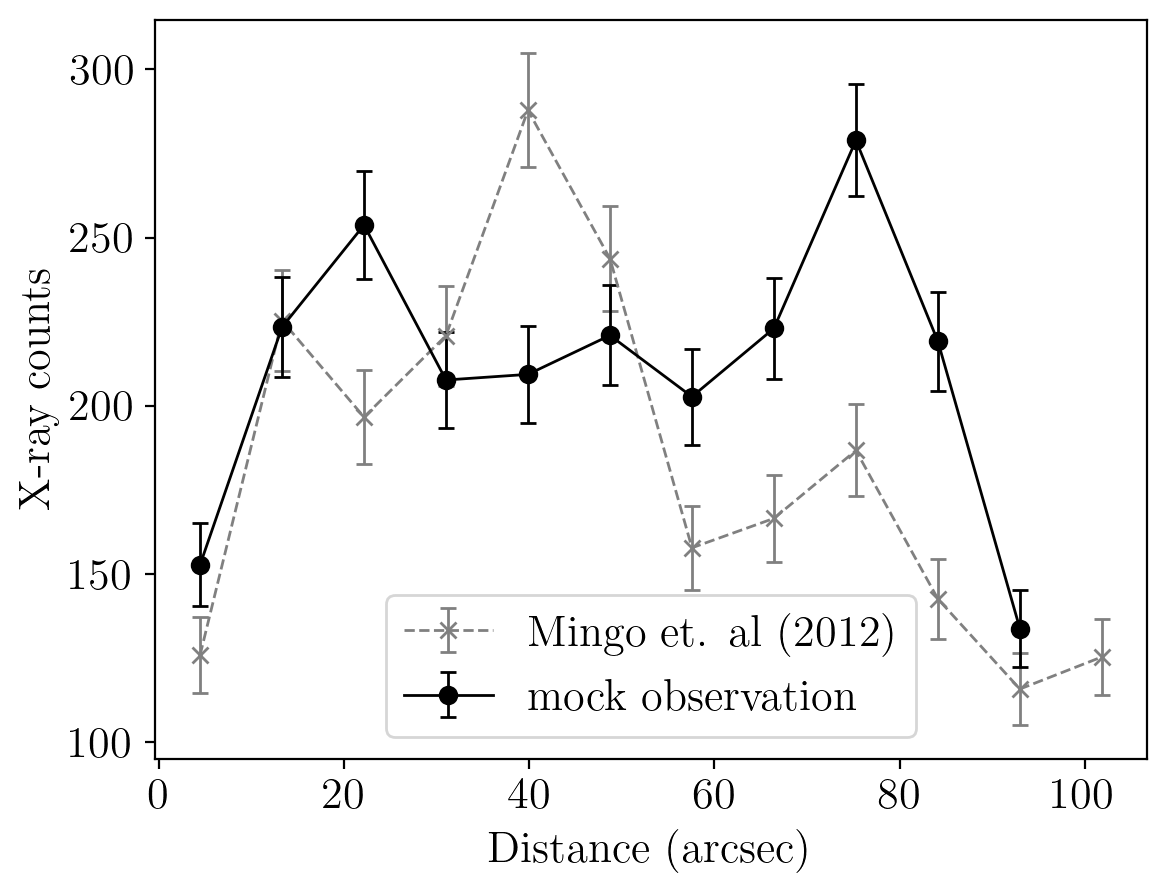}
	\caption{0.4 – 5.0 keV X-ray surface-brightness profiles along the cyan rectangular sectors marked in Fig. \ref{fig:mock_xray} (see \citetalias{mingo2012}). Dotted: the observed Chandra data. Solid: the mock X-ray observation from our fiducial jet simulation.}
	\label{fig:sb_profile}
\end{figure}

The resulting mock Chandra image in the $0.4 - 5.0\Unit{keV}$ band is shown in the right panel of Fig. \ref{fig:mock_xray}. It also includes the effects of the instrument background and the default Galactic foreground emissions from the hot Milky Way halo and the Local Bubble \citep{McCammon2002} automatically added by \texttt{SOXS}. For direct comparison, the actual Chandra observation is shown in the left panel of Fig. \ref{fig:mock_xray}. The identical green ovals in both panels roughly show the outer edges of the bubbles and have the same physical size as the dashed ellipse in Fig. \ref{fig:xray_map} after converting from the angular scale to the physical scale. In Fig. \ref{fig:sb_profile}, we show a quantitative comparison of the surface-brightness profiles along the identical cyan-segmented sectors marked in the left and right panels of Fig. \ref{fig:mock_xray} (see \citealt{mingo2012}). As expected, both the observed and synthetic bubbles are bright at the edge and fade towards the center. The central brightness excess seen in the real data, which is referred to as the ``hotspot'' region in \citetalias{mingo2012}, can not be reproduced in our hydrodynamic simulation, and may be due to additional non-thermal emission from the ejecta bubble or unresolved components not yet included in our simulation.

Potential non-thermal emissions from the hotspot region include inverse-Compton emission and synchrotron self-Compton (SSC) emission by CRes in the ejecta bubble. In inverse Compton scattering, a seed photon of energy $E_\mathrm{i}$ interacting with a cosmic-ray electron of Lorentz factor $\gamma$ is boosted to $E_\mathrm{IC}\sim\gamma^2 E_\mathrm{i}$. Taking SSC emission as an example, electrons with $\gamma \sim 2 \times 10^4$ ($\sim 10\Unit{GeV}$) can up-scatter typical $1.4\Unit{GHz}$ synchrotron seed photons to $E_\mathrm{IC}\approx 2.3\Unit{keV}$, right within the Chandra band. Furthermore, the SSC power scales as $P_\mathrm{SSC}\propto n_\mathrm{CRe}^2$, and thus becomes particularly important for regions of high CRe density. We track the evolution of the jet ejecta with a passive scaler tracer (see \citealt{duan20} for the method), and find that the jet ejecta concentrate in the hotspot region near the working surface of the jet at $t=t_\mathrm j$. This would naturally produce a relatively compact SSC X-ray enhancement in the bubble interior. 

\begin{figure}
	\plotone{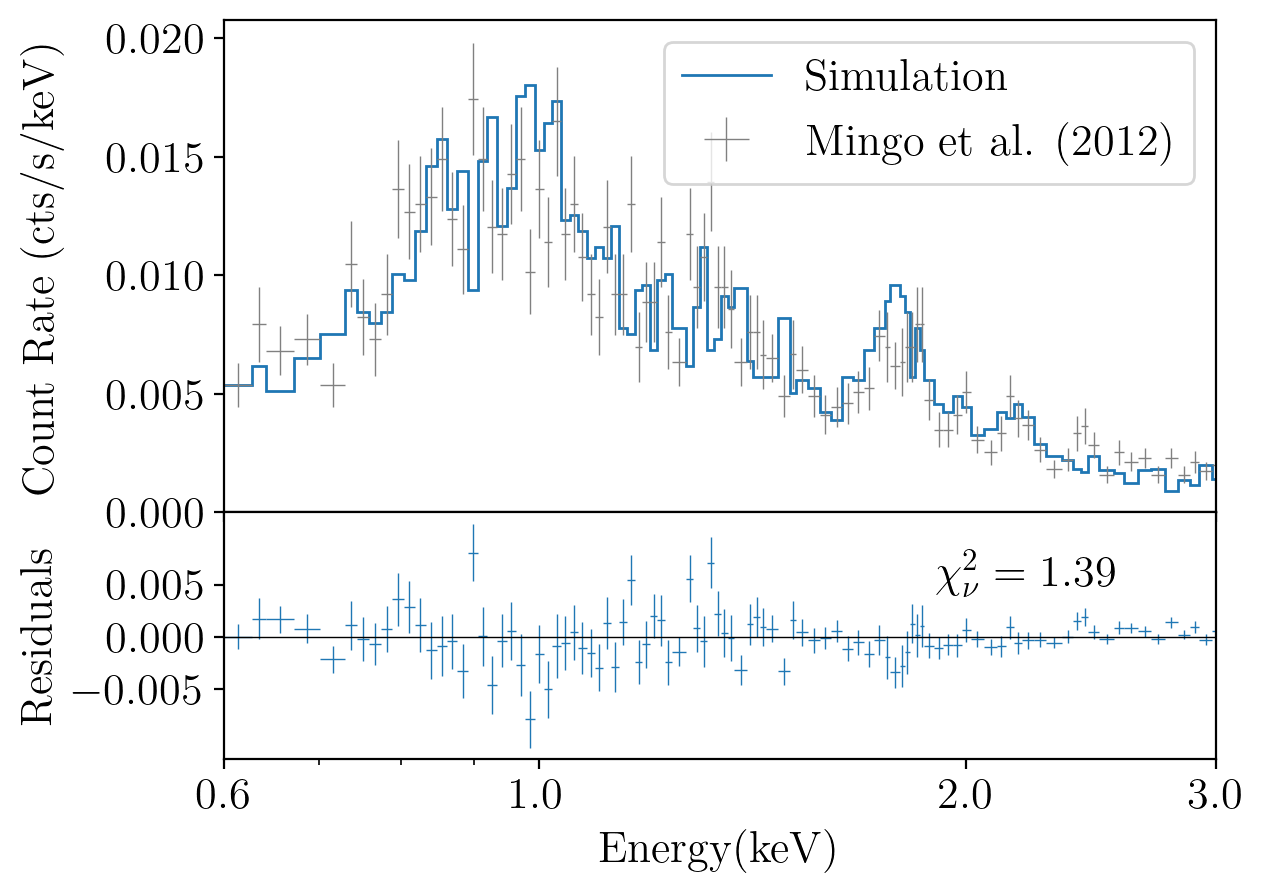}
	\caption{Comparison between the Chandra spectrum (gray data points with error bars; \citetalias{mingo2012}) and the mock X-ray spectrum (solid curve) from our jet simulation (including foreground, background and instrument response). Both are extracted from a representative region covering most of the western Circinus bubble (\citetalias{mingo2012}), marked as the identical yellow polygons in the left and right panels of Fig. \ref{fig:mock_xray}. The residuals are shown in the lower panel. The close match across the 0.6 – 3.0 keV band confirms that our jet simulation reproduces the observed X-ray emission of the western Circinus bubble quite well.}
	\label{fig:spectrum}
\end{figure}

From our mock photon event file and the Chandra data, we extracted the spectra from a representative region covering most of the western Circinus bubble (\citetalias{mingo2012}), while carefully excluding the emission from the AGN. This same region is marked as the identical yellow polygons in the left and right panels of Fig. \ref{fig:mock_xray}. Fig. \ref{fig:spectrum} compares the observed Chandra spectrum (gray data points with errors; \citetalias{mingo2012}) with our mock spectrum (solid curve); the lower panel shows the residuals. The mock spectrum, which incorporates the Galactic foreground emission and absorption, instrument background, and realistic detector response, closely matches the observed spectrum across the $0.6-3.0\Unit{keV}$ band, indicating that thermal emission from the post-shock region in our jet simulation naturally explains most of the observed X-ray emission from the western Circinus bubble.

\begin{figure}
	\plotone{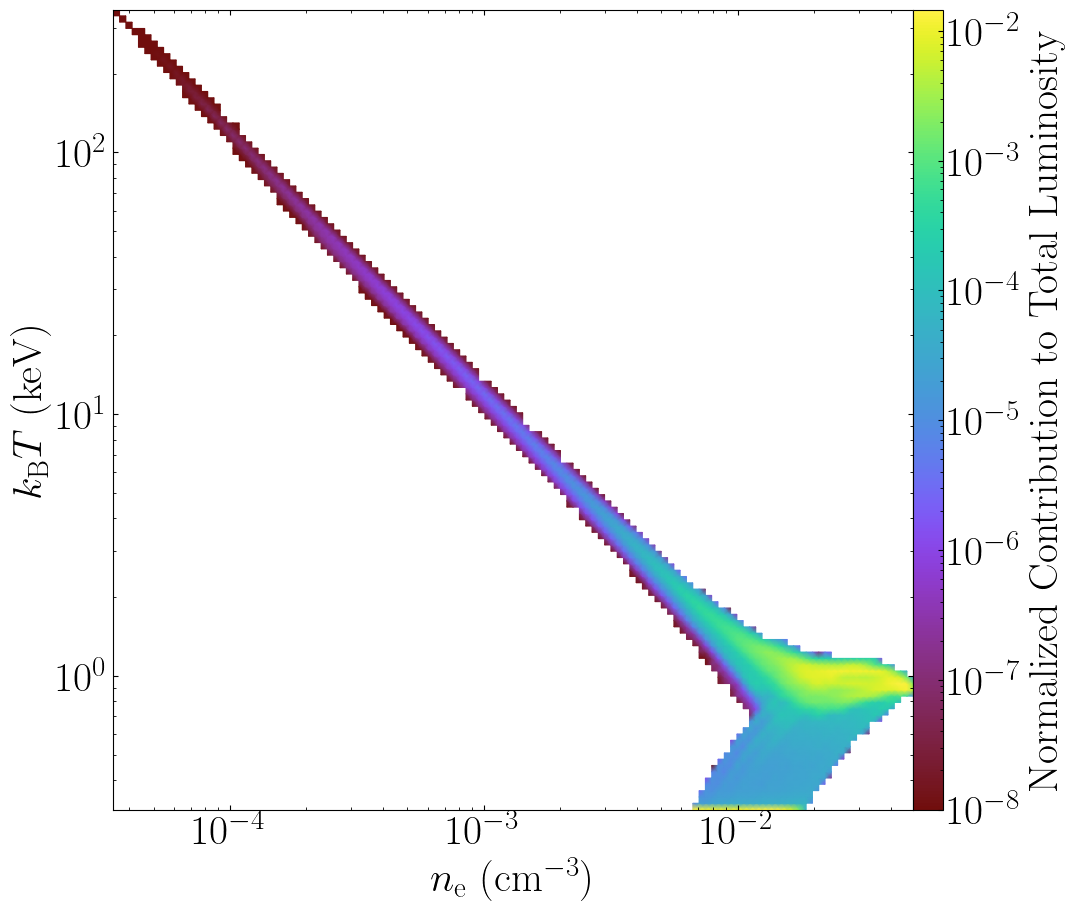}
	\caption{Temperature–density phase diagram of the gas at $z > 1\Unit{kpc}$ in the \texttt{jet} simulation. The color represents the contribution of the cells in the phase space to the total X-ray luminosity in the $0.4$ – $5.0$ keV band. It is clear that the majority of the X-ray emission arises from the hot gas in the post-shock region with $k_\mathrm{B}T \approx 0.9\Unit{keV}$ and $n_e \approx (2-5)\times10^{-2}\Unit{cm^{-3}}$.}
	\label{fig:phaseplot}
\end{figure}

We further investigate the properties of the X-ray emitting plasma in the jet simulation at $t_\mathrm{bub} \simeq 0.95\Unit{Myr}$ in Fig. \ref{fig:phaseplot}, which shows the temperature-density phase diagram of the gas at $z > 1$ kpc (to avoid contamination from the dense nuclear region).
The color represents the relative contribution of the cells in the phase space to the total X-ray luminosity in the $0.4$ – $5.0$ keV band. The intact CGM gas beyond the simulated bubble forms the coolest, lowest-density component. Above it lies the gas in the post-shock region, characterized by higher temperatures and densities. The inner jet ejecta span roughly two orders of magnitude in both temperature and density, while having almost the same gas pressure. The bulk of the X-ray luminosity arises from the gas with $k_\mathrm{B}T\sim 0.9\Unit{keV}$ and $n_\mathrm{e}\approx (2-5)\times 10^{-2}\Unit{cm}^{-3}$, which are precisely the properties of the gas in the post-shock region as shown in Fig. \ref{fig:2d}, confirming that the observed X-ray emission from the western Circinus bubble is dominated by the radiation from the post-shock halo gas. 

\section{Discussion\label{sec:discussion}}
\subsection{Dependence on Model Parameters\label{sec:parameters}}

As shown in the previous section, our representative jet run reproduces the observed properties of the western Circinus bubble quite well. We have performed a suite of jet simulations to explore the dependence on model parameters, but it is impractical to explore the full parameter space. Here, we briefly discuss how our results are affected by several main jet parameters. 

Analytical studies of jet propagation \citep{bromberg2011} show that, in the absence of strong environmental density gradients, the advance speed of the jet ``head'' is determined by its ram pressure ($\propto \eta v_\mathrm{j}^2$), while the lateral expansion is driven by the cocoon pressure. For the kinetic-energy dominated, sub-relativistic jets considered here, higher kinetic luminosity (larger $\eta$ or $v_\mathrm{j}$) results in bubbles elongated more significantly along the jet direction during its active phase, while higher internal energy density (larger $\kappa$) leads to more spherical bubbles. However, after the jet switches off, the bubble evolves toward a pressure-driven Sedov-Taylor phase, and the bubble shape gradually becomes more spherical with time. Thus, for the same jet power, a smaller $t_\mathrm{j}$ leads to a more spherical bubble with a larger age ($t_\mathrm{bub}$). This evolution is consistent with earlier numerical studies \citep{guo2015}, which found that light, fast jets tend to produce spherical ejecta bubbles with broad heads, whereas heavier, slower jets tend to generate narrower bubbles elongated along the jet direction.

Although degeneracies between some model parameters (e.g., jet density $\eta$ and velocity $v_\mathrm{j}$) cannot be easily broken by current observations, the jet power ($P_\mathrm{j}\propto \eta v_\mathrm{j}^3$) may be reasonably well constrained by the temperature of the X-ray emitting gas (mainly in the post-shock region), which has been measured by X-ray spectral observations (\citetalias{mingo2012}). A jet with higher (lower) power tends to trigger stronger (weaker) forward shocks, which heat the post-shock gas to higher (lower) temperatures. Although the value of $t_\mathrm{j}$ also affects the shock's Mach number, it can be constrained by the current location of the hotspot region and other jet parameters. In addition, the overall magnitude of the X-ray surface brightness distribution directly probes the normalization of the initial CGM density $\rho_0$. 

\subsection{The AGN Wind Model}

\begin{figure*}
\centering
\includegraphics[height=12cm]{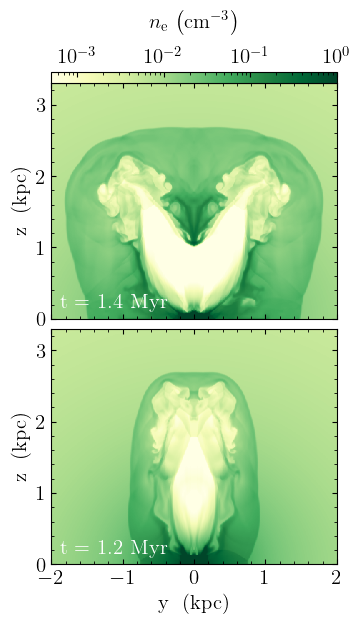}
\includegraphics[height=12cm]{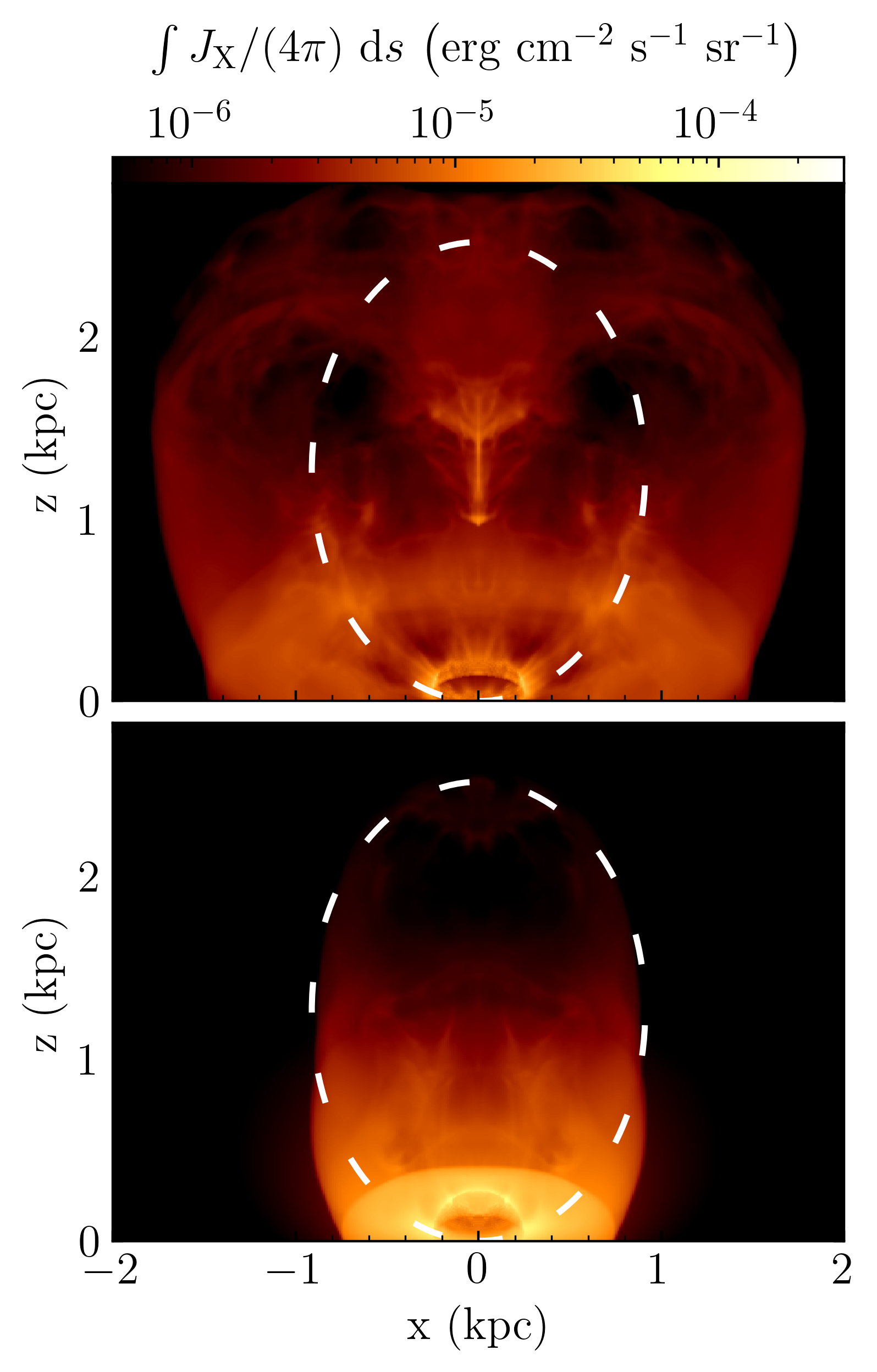}
\caption{Results of our representative AGN wind runs \texttt{wind60} (half-opening angle $60\degr$; top panels) and \texttt{wind30} (half-opening angle $30\degr$; bottom panels) at $t=t_\mathrm{bub}$ (see Table \ref{tab:runs}). Left column: the electron number density distribution in the central $x$ – $z$ plane. Right column: the synthetic 0.4 – 5.0 keV X-ray surface brightness distribution (APEC model, $Z = 0.25\;Z_\odot$). The dashed ellipses in the right panels mark the observed outer boundary of the western Circinus bubble. The bubble in run \texttt{wind60} is much more spherical than the observed one, and that in run \texttt{wind30} does not reproduce the conical shape of the bubble base. 
\label{fig:wind}}
\end{figure*}

In addition to collimated jets, wide-angle disk winds are another commonly invoked outflow mechanism in AGN feedback models \citep{yuan2014,yuan2015,yoon2019} and have been proposed as a possible mechanism to explain the origin of the Fermi/eROSITA bubbles in the Milky Way \citep{mou2014, mou2023}. In this subsection, we investigate whether such AGN winds can account for the Circinus bubbles, and adopt the analytical approximations of wind properties derived by three-dimensional general-relativisitic magnetohydrodynamic (GRMHD) simulations of hot accretion flows in \citet{yuan2015}. The properties of hot AGN disk winds remain poorly constrained in observations, but these analytical approximations have been successfully used in previous galaxy-scale hydrodynamic simulations of galaxy evolution \citep{yuan2018,yoon2019,zhu2023}.

According to \citet{yuan2015}, the wind's poloidal velocity $v_{\mathrm{wind}} \approx 0.25 v_\mathrm{k}(r_{\mathrm{tr}})$, where $v_\mathrm{k}(r_{\mathrm{tr}})$ is the Keplerian velocity at the truncation radius $r_{\mathrm{tr}}$ separating the outer thin accretion disk and the inner hot accretion flow. The value of $v_{\mathrm{wind}}$ remains nearly constant on a relatively large scale throughout their entire simulation domain. The wind is mainly distributed within a half opening angle of $\sim 60^\circ$, with faster outflows generated in the central region within a half opening angle of $30^\circ$. As shown in \citet{yuan2014}, higher accretion rates push the truncation radius $r_{\mathrm{tr}}$ inward, resulting in faster winds. For a pure hot accretion flow, the maximum accretion rate is $\sim 0.02 \dot M_\mathrm{Edd}$, implying an upper limit of $v_\mathrm{wind} \approx 0.1c$ for the $1.7\times 10^6 M_\odot$ supermassive black hole of Circinus.

The wind implementation in our simulations is essentially the same as our jet implementation method described in Section \ref{sec:jet}, except that the central cylindrical jet nozzle is replaced by a spherical cone with radius 100 pc. The apex of this cone is located at the galactic center, and we inject the wind with a constant mechanical power. Here, we present the results of two representative wind simulations, run \texttt{wind60} and run \texttt{wind30}, for AGN winds with half-opening angles of $60\degr$ and $30\degr$, respectively. The wind properties in these two runs are listed in Table \ref{tab:runs}.

Fig. \ref{fig:wind} shows the results of these two wind runs at the final epoch $t=t_\mathrm{bub}$ when the forward shock reaches $z=2.7\Unit{kpc}$ (see Table \ref{tab:runs} for the specific values of $t_\mathrm{bub}$). The left panels show the electron number density distribution in the central $x$ – $z$ plane, and the right panels show the corresponding synthetic 0.4 – 5.0 keV X-ray surface brightness distribution. The dense wind ejecta produce substantial X-ray emission from the bubble interior, substantially diminishing the observed shell-to-interior contrast in X-ray surface brightness. Furthermore, the wide-angle wind in run \texttt{wind60} produces a bubble with a poorly fitting aspect ratio, which is much more spherical than the elongated shape of the observed bubble. The narrowing of the half-opening angle to $30\degr$ in run \texttt{wind30} improves the match of the bubble aspect ratio, but the base of the simulated bubble is still too broad and does not reproduce the observed conical shape. This result is consistent with previous studies of \cite{zhang2020}, who find that spherical winds tend to produce bubbles with bases much wider than the observed Fermi bubbles. A nuclear gaseous disk (e.g., the central molecular zone in the Milky Way) may suppress the lateral motion of the wind ejecta near the galactic plane, but may still be ineffective in suppressing the lateral propagation of the forward shock. 

\begin{figure}
	\plotone{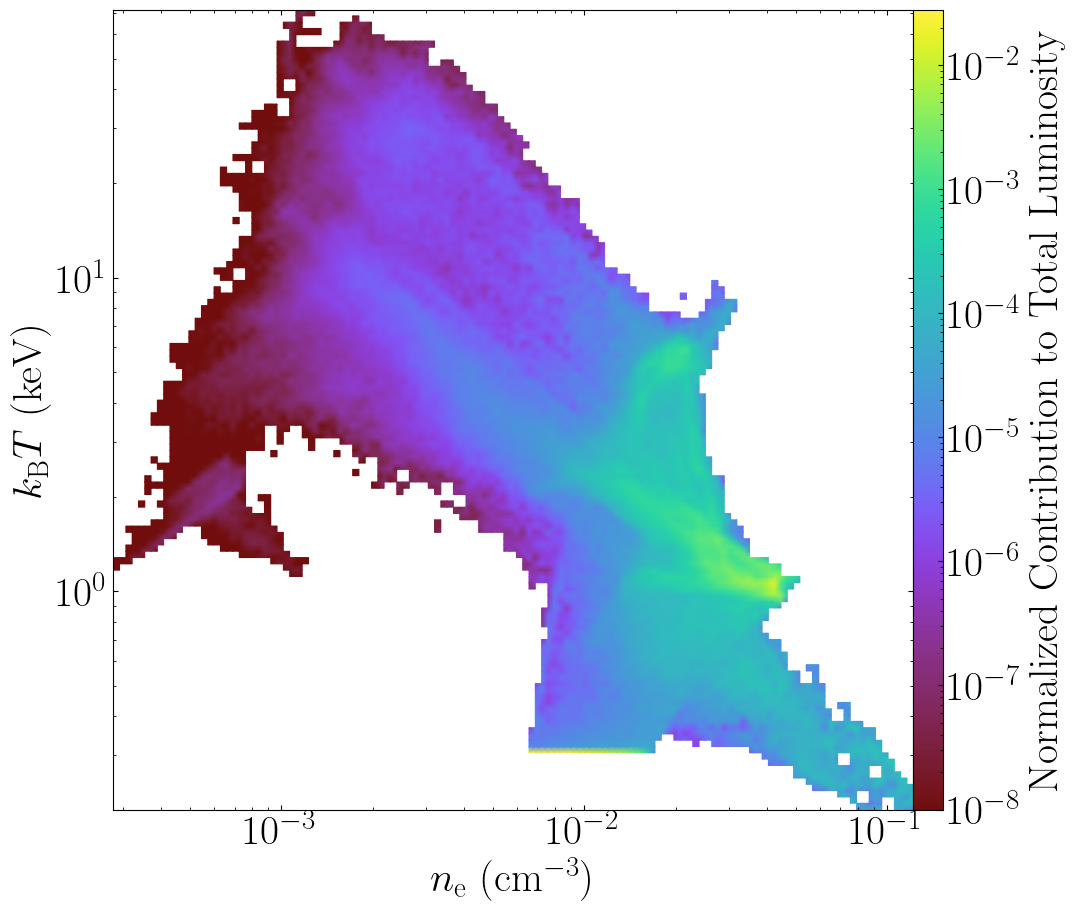}
	\caption{Temperature–density phase diagram of the gas at $z > 1\Unit{kpc}$ in the \texttt{wind30} simulation. The color represents the contribution of the cells in the phase space to the total X-ray luminosity in the $0.4$ – $5.0$ keV band. The X-ray emission is contributed mainly by the hot gas with temperatures spanning more than one order of magnitude, inconsistent with the single-temperature spectral fit to X-ray observations of the western Circinus bubble (\citetalias{mingo2012}).}
	\label{fig:phaseplot_wind}
\end{figure}

Fig. \ref{fig:phaseplot_wind} shows the temperature–density phase diagram of the gas at $z > 1\Unit{kpc}$ in run \texttt{wind30}, and the color here represents the contribution of the cells in the phase space to the total X-ray luminosity in the $0.4$ – $5.0$ keV band. In stark contrast to the jet-driven bubble (Fig. \ref{fig:phaseplot}), where the X-ray emission is dominated by the gas at a nearly single temperature ($k_\mathrm{B}T \approx 0.9\Unit{keV}$), the X-ray emission from the wind-driven bubble is contributed mainly by the hot gas with temperatures spanning more than one order of magnitude. This broad temperature distribution is inconsistent with the single-temperature spectral fit to X-ray observations of the western Circinus bubble (\citetalias{mingo2012}).

The broad temperature distribution of the post-shock gas in wind simulations is directly related to the relatively high value of the wind duration $t_\mathrm j$ (see Table \ref{tab:runs}). Wide-angle winds tend to produce more spherical bubbles than collimated jets, and bubble evolution after wind cessation renders the bubble even more spherical. Therefore, to produce the radially-elongated shape of the Circinus bubbles, $t_\mathrm j$ in wind simulations must be large enough to minimize the bubble evolution after the wind cessation. When the wind is active, the forward shock is strongest along the $z$ axis and weakens further away from this axis, resulting in substantial temperature variation in the post-shock gas. However, the jet in run \texttt{jet} is switched off much earlier, allowing substantial bubble evolution in the later pressure-driven “Sedov-Taylor” phase. During this phase, the gas pressure in the jet ejecta is nearly uniform, driving shocks of a comparable Mach number ($\sim 2$ - $3$ at $t=t_\mathrm{bub}$) in all directions, and consequently the post-shock gas temperature becomes approximately uniform.

\subsection{Comparison between Fermi Bubbles and Circinus Bubbles}

The Circinus bubbles share many key similarities with the Fermi bubbles in the Milky Way. Both are kpc-scale elliptical bubbles containing extended diffuse emissions with sharp edges detected in a spiral galaxy. Similar to the Circinus bubbles, Fermi bubbles have also been observed in X-rays \citep{bh03,zhang2020} and radio (WMAP observations; \citealt{finkbeiner2004}, \citealt{dobler2008}). Our AGN jet-shock model has successfully explained the key observations of both the Fermi bubbles \citep{zhang2020,zhang25} and the Circinus bubbles (this work). Therefore, it is quite likely that both belong to the same kind of AGN jet feedback occurring in regular disk galaxies.

Our simulations also show that the Circinus bubbles are significantly younger ($t_\mathrm{bub}\sim 1\Unit{Myr}$) than Fermi bubbles ($t_\mathrm{bub} \sim 5\Unit{Myr}$; \citealt{zhang2020}). This implies that the cosmic rays accelerated in the former are much younger than those in the latter. In particular, the relatively old age of the Fermi bubbles explains why the jet ejecta (the ``inner bubble'') are not bright enough to be unambiguously detected in the gamma-ray band \citep{ackermann14}. The slightly old CRes in the jet ejecta could still emit synchrotron radiation and may have already been observed in the microwave band \citep{del24,zhang24}. In contrast, the Circinus bubbles are much younger and clearly contain an inner ejecta-filled bubble (the hotspot region) that is emitting in detectable X-rays and radio (\citetalias{mingo2012}). Gamma-ray emissions in the GeV band from Circinus have also been detected \citep{hayashida13,guo2019}. Compared to the central starburst region, AGN activity may contribute sub-dominantly to gamma ray emission, but it is naturally expected from their low age that both the Circinus bubbles and the inner ejecta bubbles produce substantial $\gamma$-ray radiation, which may be observed by future $\gamma$-ray observing facilities with sensitivity and angular resolution much higher than current ones. 

\section{summary}
\label{sec:summary}

The nearby spiral galaxy Circinus hosts a pair of kpc-sized elliptical bubbles observed in the X-ray and radio bands. These diffuse bubbles extend into the galactic halo and are strongly reminiscent of Fermi bubbles in the Milky Way. The Circinus bubbles have a slightly smaller size and substantial hotspot emissions, suggesting that they may be potentially in a younger evolution stage than the Fermi bubbles.

In this paper, we used 3D hydrodynamic simulations to investigate the formation of Circinus bubbles in the modeled gravitational potential and the CGM of Circinus. We find that a pair of AGN jets drive forward shocks in the CGM, and after evolving for $\sim 0.95\Unit{Myr}$, the shock-delineated CGM bubble roughly matches the observed western Circinus bubble in size and morphology. The jet in our fiducial \texttt{jet} run is light, over-pressured relative to the ambient medium, and kinetic-energy-dominated with a total power of $P_\mathrm{j}=1.5\times 10^{42}\Unit{erg/s}$ for the jet pair. The duration of the jet is $0.5\Unit{Myr}$ and the current age of the Circinus bubbles is predicted to be $t_{\text{bub}}\sim 0.95\Unit{Myr}$.

Our synthetic 0.4 – 5.0 keV X-ray image reproduces the observed edge-brightened X-ray surface brightness distribution quite well, and suggests that non-thermal emissions from the jet ejecta also contribute substantially to radio and X-ray emissions from the bubble's hotspot region. Our mock Chandra image and spectrum also fit the real Chandra data quite well, indicating that the X-ray emission of the bubbles is mainly contributed by the post-shock region with density $n_\mathrm{e} \sim 0.1$ – $1 \Unit{cm^{-3}}$ and temperature $k_\mathrm{B}T \approx 0.9\Unit{keV}$.

We further show that AGN winds tend to produce more spherical bubbles with a wider base near the galactic plane, which is inconsistent with the observed bubble morphology elongated along the jet direction with a conical base. AGN winds also tend to produce a broad temperature distribution for the post-shock gas, which is inconsistent with the single-temperature fit to the Chandra X-ray spectrum of Circinus bubbles (\citetalias{mingo2012}). In the Introduction (Section \ref{sec:intro}), we also argued that the starburst wind model is disfavored.

The AGN jet-shock model has previously been invoked to successfully explain the observations of Fermi bubbles \citep{zhang2020,zhang25}. Our study thus further corroborates this model as the origin of both the Circinus bubbles and the Fermi bubbles, and suggests that in addition to starburst winds, AGN jet feedback may be another common origin of extended gaseous bubbles in regular disk galaxies, potentially playing an important role in their evolution. In our model, the Circinus bubbles are significantly younger ($t_\mathrm{bub}\sim 1\Unit{Myr}$) than the Fermi bubbles ($t_\mathrm{bub} \sim 5\Unit{Myr}$), implying that the CRes transported by the jet ejecta are much younger and potentially account for the enhanced X-ray and radio emissions of the hotspot region (the ``ejecta bubble''). They may even produce substantial gamma ray emission via inverse Compton scattering of low-energy photons in the cosmic microwave background and the interstellar radiation field, in contrast with the much-older ejecta bubble of Fermi bubbles, which has not been unambiguously detected in gamma rays \citep{ackermann14}.

\begin{acknowledgements}
The authors thank Fabrizio Nicastro for helpful discussions. This work was supported by the National SKA Program of China (No. 2025SKA0130100), the Excellent Youth Team Project of the Chinese Academy of Sciences (No. YSBR-061), the National Natural Science Foundation of China (No. 12473010), the Shanghai Pilot Program for Basic Research - Chinese Academy of Sciences, Shanghai Branch (No. JCYJ-SHFY-2021-013), and the China Manned Space Program (Nos. CMS-CSST-2025-A08 and CMS-CSST-2025-A10). BM acknowledges support from UKRI STFC for an Ernest Rutherford Fellowship (grant no. ST/Z510257/1). J.-T. L. acknowledges the financial support from the National Science Foundation of China through the grants 12321003 and 12273111, the China Manned Space Program with grant no. CMS-CSST-2025-A04 and CMS-CSST-2025-A10, and Jiangsu Innovation and Entrepreneurship Talent Team Program through the grant JSSCTD202436. The simulations presented in this work were performed using high-performance computing resources in the Core Facility for Advanced Research Computing at Shanghai Astronomical Observatory.
\end{acknowledgements}

\bibliography{refs}
\bibliographystyle{aasjournal}
\end{document}